\documentclass[11pt,twoside]{article}
\usepackage{newpasp}
\usepackage{psfig}
\usepackage{natbib}
\usepackage{epsf}
\usepackage{graphics}
\usepackage{lscape}
\def\edcomment#1{\iffalse\marginpar{\raggedright\sl#1\/}\else\relax\fi}
\reversemarginpar

\begin{document}
\title{The COMPLETE Survey of Star-Forming Regions on its Second Birthday}
\author{Alyssa A. Goodman\altaffilmark{1}}
\altaffiltext{1}{AG writes on behalf of the full COMPLETE consortium, which includes members at: the CfA, USA (Michelle Borkin, Jonathan Foster, Di Li,  Naomi Ridge, and Scott Schnee); ESO, Germany (Jo\~ao Alves and Tom Wilson); Caltech, USA (H\'ector Arce); Arcetri, Italy (Paola Caselli);  HIA, Canada (James DiFrancesco, Doug Johnstone, and Helen Kirk); UMASS/FCRAO, USA (Mark Heyer); and OAN, Spain (Mario Tafalla).}
\affil{Harvard-Smithsonian Center for Astrophysics}

\begin{abstract}

At age two, humans are just learning to speak, and they see the world in an unbiased way.  At age two, the COMPLETE ({\bf CO}ordinated {\bf M}olecular {\bf P}robe {\bf L}ine {\bf E}xtinction {\bf T}hermal {\bf E}mission ) Survey of Star-Forming Regions is just beginning to hint at what it will do for our world view, and it is unbiased as well.  The coordinated observations that comprise COMPLETE  have already shown us: 1) that large ($\sim 10$ pc) scale bubbles created by winds from evolved stars dominate maps of extended dust emission even in relatively low-mass star-forming regions (e.g. Perseus and  Ophiuchus); 2) that ``giant" ($>1$ pc-scale) outflows from young stars can be easily detected by applying statistical tools to an unbiased CO survey (e.g. we have doubled the number of known outflows in Perseus using the Spectral Correlation Function); and 3) that many apparent high column-density ``clouds" seen in projection are actually the superposition of many velocity features (likely at different distances) along the line of sight.  The upcoming detailed analysis of these new observations will quantify the degree to which the apparently ubiquitous spherical and collimated outflows can sculpt and stir molecular clouds.    Taken together with recent numerical simulations' results and measurements of high velocities for some young stars, the COMPLETE Survey also raises, and should soon answer, questions about how, when, and how far stars move away from their birthplaces as functions of time.

\end{abstract}

\section{Introduction}
The prevailing theoretical picture of star formation envisions stars forming inside of dense cores, which are in turn embedded in larger, slightly lower-density structures.  Each forming star is surrounded by a disk, and, when it is very young, the star-disk system produces a collimated bipolar flow, in a direction perpendicular to the disk.  In its broad outlines, this paradigm is very likely to be right.  In detail, though, many questions concerning the timing of this series of events remain.  For example, how long does a star stay with its natal core?  How long does it remain associated with the Òlower-density structureÓ (e.g. a filament in a dark cloud) where it originally formed?  What kind of environment does a star-disk system need to keep accreting, or to produce an outflow--and when might that reservoir no longer be available to the system?   Does a bipolar outflow have any effect on star formation nearby? What causes fragmentation into binaries or higher-order systems?  How much influence do spherical winds  (e.g. SNe, B-star winds) from previous generations of stars have on the timing of star formation?  How often is a star in the process of forming likely to encounter an external gravitational potential (e.g. from another forming star) strong enough to alter its formation process?
The complicating issue underlying all of these questions is that it is hard to define and understand the long-lasting properties of the reservoir from which a star forms if the reservoir itself is highly dynamic.

The COMPLETE Survey of Star-Forming Regions was born out of a desire to create an unprecedented comprehensive statistical database with which one might have real hope of answering many of these questions.   We were inspired to begin COMPLETE at the 2001 Santa Cruz Star Formation Workshop, while grousing at a coffee break about how virtually none of the beautiful measurements of cloud's velocity fields (from molecular line observations), density profiles (from extinction measurements),  temperature and dust property profiles (from thermal emission mapping), and embedded source distributions (from infrared imaging) we had been seeing in the talks covered the {\it same} region of space.   In other words, since people usually observed their ``favorite" object(s) with their favorite technique, we almost never knew then density, temperature, and velocity field of the same exact set of objects.\footnote{Elizabeth Lada's pioneering thesis work \citep{1992ApJ...393L..25L, 1991ApJ...371..171L, 1991ApJ...368..432L}  provides a notable exception to this trend.  Through years of work, Lada and colleagues mapped the actively star-forming cloud L1630 (a.k.a. Orion B)  in both molecular lines (CS, with 2$'$ resolution) and with near-infrared cameras (reaching $m_K<13$ mag).  Lada et al.'s work provided the first evidence that massive stars form in clusters, and it also showed that the mass spectrum of the gaseous material (self-gravitating or not) is shallower than that of stars.  COMPLETE, which was not feasible a decade ago,  should do for low-mass (fainter) star-forming regions what Lada's work did for (brighter) massive star-forming regions, and more.  For comparison, the total areal coverage of COMPLETE is $\sim 20$ square degrees, which is an order of magnitude larger than the area Lada et al. studied in Orion.}    

During that same Santa Cruz meeting, we calculated that a  ``{\bf CO}ordinated {\bf M}olecular {\bf P}robe {\bf L}ine {\bf E}xtinction {\bf T}hermal {\bf E}mission" (COMPLETE) Survey covering tens of square degrees of sky would take only many weeks of observing time, rather than the preposterous many decades of time it would have taken 15 years earlier.   It was obvious that if we could  create such a database, and make it publicly available, then many researchers (including us!) could finally combine all of the powerful techniques developed over the past twenty years, in order to measure the physical properties of a carefully chosen set of star-forming regions.  It was also obvious that we should chose regions that were then about-to-be, and now are in the midst of being, observed by the Spitzer Space Telescope under the ``Cores-to-Disks" (c2d) Legacy Program.\footnote{More information about c2d and about COMPLETE (including online data) are available through the COMPLETE web site, at cfa-www.harvard.edu/COMPLETE.}  By choosing the three Northern Hemisphere c2d extended cloud targets (Perseus, Ophiuchus and Serpens), we would be assured that a  full census of the embedded stellar population and its properties would be available around the same time that COMPLETE was finished.  

Today, the ``COMPLETE"  Survey conceived at the infamous Santa Cruz coffee break is a healthy two-year old large international collaboration.$^1$  In the winter of 2003/4, we have reached the halfway point in COMPLETE's observations, and  Spitzer is up and about to begin pouring out the c2d Legacy data. 

\section{COMPLETE Observations}

COMPLETE was designed to be executed in two phases.  Phase 1, from 2002-2004 focuses on observing the ``larger context" of the star formation process (on the 0.1 to 10 pc scale).   {\it All} of the Perseus, Ophiuchus and Serpens cloud area to be observed with Spitzer under c2d, and slightly more, will be covered by COMPLETE in an unbiased way in Phase 1.   Phase 2, from 2004-2006, is a statistical study of the small-scale picture of star formation, aimed at assessing the meaning of the variety of physical conditions observed in star forming cores (on the $<0.1$ pc scale).  In Phase 2, targeted source lists based on the Phase 1 data are being used, as it is (still) not feasible to cover {\it every} dense star-forming peak at high-resolution.  

Phase 1 observations (full coverage of all three extended clouds)  include: FCRAO  mapping (40$''$, $0.06$ km s$^{-1}$ resolution) of the molecular gas (${^{12}}$CO and ${^{13}}$CO); 2MASS-based extinction mapping ($\sim 3'$ resolution) using the NICER technique \citep[see][and references therein]{2001A&A...377.1023L}; re-processing of the 60 and 100 $\mu$m ISSA plates to create new maps of dust column density and temperature; and SCUBA/JCMT mapping at 850 $\mu$m (14$''$ resolution)  of all regions where $A_V>5$. To give a feel for the enormity of this data set, we can point out that just one of our molecular line maps of Perseus has $\sim 10^5$ spectra of $10^3$ channels each.  The entire data set is freely available at cfa-www.harvard.edu/COMPLETE.

Phase 2, which has just begun, includes IRAM 30-m spectral line and continuum (MAMBO) mapping of selected cores (resolution $\sim 10-20''$), and higher-resolution extinction mapping using both the c2d data itself and ground-based near-IR data from 4- to 8-m class telescopes.

%\begin{figure}[!ht]
%\plotone {overlay.eps}
%\plotone {goodmanfig1.eps}
%\end{figure}

\section{Warm Bubbles in ``Cold" Clouds?}

Figure 1 shows COMPLETE's velocity-integrated $^{13}$CO map\footnote{Figure 1 is identical to an image shown at the meeting summarized in this volume.  Current maps have more coverage, and are available at cfa-www.harvard.edu/COMPLETE.} of Perseus  overlaid on a new  dust column density map of the region created using 60 and 100 $\mu$m ISSA data\footnote {See \citet{1999ApJ...512L.135A}  and \citet{ridge04} for methodology.}.  The molecular line map is dominated by the long chain of clouds first mapped out by \citet{1986A&A...166..283B}, while the ISSA (IRAS) based map clearly shows a big bubble.\footnote{The 2MASS/NICER extinction map, available at the COMPLETE web site, looks morphologically like the integrated $^{13}$CO map.}    Figure 2 shows that the ring dominating the 60/100-$\mu$m-based column density map of Perseus is apparently heated by even warmer material interior to it.  

This kind of morphological dissimilarity between warm dust and cold gas is also obvious in the famous IRAS image of the Orion molecular clouds, which is dominated by large shells emanating from evolved stars \citep[e.g.][]{1989A&A...218..231Z}.  In low-mass star forming regions like Taurus, however, an IRAS-based map is nearly indistinguishable from an integrated $^{13}$CO map or an extinction map \citep{1994ApJ...423L..59A}.  In regions exhibiting a mixture of warm and cool dust, a 60/100-$\mu$m-derived column density map can bear a strong resemblance to the cool clouds' morphology, but the dust temperature map often highlights shells.  For example, our COMPLETE ISSA-based dust temperature map of Ophiuhcus (available online) shows a 2-pc diameter heated ring $\sim$around $\rho$-Oph, centered a degree north of the so-called ``$\rho$-Oph" cluster \citep[see][]{li04}).  
%\begin{figure}[!ht]
%\plotone {perseus_iras.eps}
%\plotone {goodmanfig2.eps}
%\end{figure}

In general, how similar an extinction or molecular-line map looks to a dust emission map depends sensitively on the mixture of dust temperature and composition along the line of sight.  At a given wavelength, the same amount of dust at a higher temperature will always emit more than it would at a lower temperature.   Thus, once significant amounts ($\geq$ amount of cold dust) of heated dust are present in any region, the appearance of that region will be dominated by the distribution of that dust in thermal emission maps.  Changes in dust composition or size distribution that are correlated with dust heating complicates the picture even further and are discussed in \citet{ridge04}.
 
\subsection{Impact Assesment}\label{assessing}

How much do warm bubbles impact the star formation process overall?  In Perseus, it is not even clear that the bubble and the molecular clouds lie at the same distance \citep[see][]{ridge04}.  On the other hand, in Ophiuchus, there is clear evidence for interaction of the warm bubble around $\rho$-Oph with the molecular gas, and it even appears that this bubble may have triggered the formation of main star-forming cluster there \citep[confusingly known as the ``$\rho$-Oph" cluster but actually a degree to the South of the star, see][]{li04}.

Simple calculations can show that the energy associated with the shells in Perseus and Ophiuchus is enough for either of them to have been generated by a $\sim 10^{51}$ erg supernova going off in dense gas, but our initial investigations favor their creation by persistent winds from evolved stars.  Whatever their origin, though, the energy in these shells may well contribute very substantially to the driving of turbulence in the dense ISM, as many theoretical simulations demand \citep[see][and references therein]{2004RvMP...76..125M}. 

When our Phase 1 molecular-line maps of Perseus (2004) and Ophiuchus (2005) are finished, we will measure the kinematic impact on the cold gas of the shells we see in warm dust directly, by analyzing  the temperature, density {\it and} velocity structure of the regions most likely to be interacting.  Figure 3 shows correlations amongst 
 2MASS(NICER)-, IRAS(ISSA)-, and $^{13}$CO-derived measures of column density in Perseus, at positions coincident with the warm ring (white dots) and with the surrounding cold gas (black dots; see Figures 1 and 2).  Notice, in Figure 3b, what a poor tracer of cold material the ISSA map is.  Notice, in Figure 3a,  how W($^{13}$CO)  ($\propto$ gas column density) is systematically higher in the cold clouds than in the ring for any given 2MASS-derived column density, even though the $N_{2MASS}$ and $W$($^{13}$CO) are well-correlated overall.   In order to correctly assess the physical influence of bubble-creating winds on the evolution of star formation in molecular clouds, this kind of confounding, confusing but ``COordinated" Molecular Probe Line Extinction Thermal Emission information is, alas, all {\it required}.  It is not possible  to extract relevant measures of the mass, temperature, and velocity of the material involved in shell/cloud interactions from any one technique alone.
%\begin{figure}[!ht]
%\plottwo{wco_n2mass.eps}{iras_2mass.eps}
%\plottwo{goodmanfig3.eps}{goodmanfig4.eps}
%\end{figure}
\section{Bipolar Outflows}\label{bipolar}
The long-term influence of bipolar outflows from young stars on star-forming regions depends on the spatial and temporal frequency of their occurrence and on the total amount \citep {1988ApJ...333..316M} and variation \citep{2001ApJ...551L.171A} of outflow energy with time. We seek to measure all of these parameters for the three regions under study in COMPLETE.  

In 2003, we undertook a systematic search for outflows in Perseus, using an early version of the huge ($>150,000$ spectra), Nyquist sampled, $^{12}$CO map being created under Phase 1 of COMPLETE \citep{2003AAS...203.7702F, ridgeoutflow04}.  By using the Spectral Correlation Function  \citep [hereafter SCF,][]{1999ApJ...524..887R, 2003ApJ...588..881P}, to identify extended patches of either very high or very low spectral correlation in the $^{12}$CO map, we can recover all ofl the known outflows in Perseus, as well as many more as ``candidate" flows.  Following up with customized channel maps of the candidate regions has allowed us, to date, to roughly double the number of known young stellar outflows in Perseus.

The task that stands before us now is use the power COMPLETE gives us to measure density, velocity, and temperature independently (see \S \ref{assessing}) in order to correctly determine the energetic impact of bipolar flows.  Unlike the case with spherical shells, where we would need nearly a galaxy's worth to do a proper statistical job, a cloud as big as one of our COMPLETE targets (e.g. Perseus) is probably large enough to assess the long-term statistical impact of the bipolar flows.  ``All" we need to do is to estimate the flows' sculpting and turbulence-driving potential now and in the past!  We will estimate outflow-driving source ages using Spectral Energy Distributions (based on combining ROSAT, 2MASS, c2d, BOLOCAM \citep{2003AAS...203.9806E}, and COMPLETE/SCUBA data), allowing us to study outflow behavior as a function of stellar age.  Finally, we will construct Monte Carlo models of outflow-cloud interactions where the temporal and spatial distribution of outflows, and the density/temperature/velocity distribution of the ambient material are all drawn from observations of a single star-forming region (e.g. Perseus).

\section{Dense Cores}

Several groups working on numerical simulations of turbulence in the ISM have pointed out that often what looks like a ``cloud" or ``core" in 2D is really just a chance column-density peak, created in 3D by the chance superposition of many physically unrelated structures along the line of sight \citep [c.f.][]{2001ApJ...546..980O}.  In scouring the COMPLETE maps of Perseus in our search for outflows (see \S \ref{bipolar}), the SCF turned up several regions which appear as column density peaks, but are in fact comprised of many different velocity components along the line of sight.  COMPLETE's high velocity resolution allows us to separate these velocity features where they would have been blended before.

Once Phase 1 of COMPLETE is done, we will calculate the fraction of column density peaks attributable to these chance coincidences (based on their spectra), rather than real 3D density enhancements. In a related effort,  we will study the spatial dependence of line width on position for the ``real" (3D) cores and for the ``fake" (2D projected) cores, in order to measure how common ``coherence'' (plateaus of near-thermal line width) is in each group \citep [see][]{1998ApJ...504..223G}.  We will also compare the distribution of  ``clumps" being found with SCUBA at 850 $\mu$m (described in Johnstone et al. 2004) with the cores identified in the spectral-line analysis.

Using Phase 1 and 2 COMPLETE data, a paper by graduate student Scott Schnee et al. will address the context of core ``rotation," by analyzing velocity gradients and spectral correlation in and around dense cores, rather than just in the cores  \citep{1993ApJ...406..528G}.    Paola Caselli and Mario Tafalla will lead an investigation of whether chemical depletion \citep [e.g.][]{1999ApJ...523L.165C, 2002ApJ...569..815T, 2004A&A...416..191T}   can be used as a clock in  dense cores.  The high-resolution NICER-based infrared extinction maps currently being proposed and constructed by Jo\~ao Alves et al. will be used as the unbiased tracer of column density in many of these core studies.

\section{What if stars move far enough from home to matter?}

Recently, in work not related to COMPLETE, we have determined that PV Ceph, a Herbig Ae/Be star, is currently moving at $\sim 20$ km s$^{-1}$, and was likely ejected, 500,000 years ago, from a cluster (NGC7023) 10 pc away from the molecular cloud where it finds itself now \citep{2004astro.ph..1486G} .  Other researchers have found that stars (still) in clusters commonly whiz around at 10's of km s$^{-1}$ \citep[e.g.][] {1995ApJ...455L.189P, 2001...546..299}.   And, numerical simulations predict that anytime stars form in groups--even small groups--some are ejected at speeds far in excess of the background velocity distribution of the gas \citep [e.g.][]{1998MNRAS.301..759K, 2002MNRAS.332L..65B} .  Therefore, it is reasonable to worry that stars might move so far beyond the confines of the core where they first formed that it would be hard to find all young stars' true birthplaces.  (For reference, at just 2 km s$^{-1}$ relative to the gas, a star will move from the center to the edge of an 0.1 pc-diameter core in just 250,000 years.)   It is possible that cores move along with the stars they form, relative to lower-density material, making star-birthplace pairing less of a problem, but it is doubtful that we will ever be able to associate every young star we see with its exact birthplace.

Even the mild motion of stars with respect to their ancestral homes will confound simplistic calculations of star formation efficiency.  So we\footnote{This effort is being led by graduate student Jonathan Foster, at the CfA.} have begun a study of star-gas motion aimed at assessing the importance of ``stellar scrambling" in a statistical way.  Ultimately, we will use COMPLETE data to map out the density and velocity distribution of gas, and existing  and new radial velocity and proper motion data to measure stellar velocities.  To estimate stellar ages, we will use the same kind of SED analysis discussed in \S \ref{bipolar}.    When the scrambling study is finished, we will be able to assess how far off an estimate of ``star formation efficiency" based on dividing the mass in young stars by the mass of star-forming gas {\it in a particular region} on the sky is likely to be.

\section{COMPLETEing the Work of the Three Wise Men}

The  {\bf CO}ordinated {\bf M}olecular {\bf P}robe {\bf L}ine {\bf E}xtinction {\bf T}hermal {\bf E}mission (COMPLETE) Survey of Star-Forming Regions was begun two years before the three wise men (Hollenbach, McKee \& Shu) turned 60, and it was explicitly designed to answer many of the questions those wise men's work have raised but not answered.  Perhaps a seventieth-birthday party gift will include some answers.

\end{document}